\documentclass[11pt,twoside]{article}

\usepackage{asp2006}
\usepackage{epsf}
\usepackage{epsfig}
\usepackage{lscape}
\usepackage{natbib}

\begin{document}
\title{Kinematic and Structural Evolution of Field and Cluster Spiral Galaxies}
\author{B. L. Ziegler\altaffilmark{1},
E. Kutdemir\altaffilmark{2},
C. Da Rocha\altaffilmark{1},
A. B\"ohm\altaffilmark{3},
R. F. Peletier\altaffilmark{2},
M. Verdugo\altaffilmark{4}}
\altaffiltext{1}{European Southern Observatory, Karl-Schwarzschild Str. 2, 85748 Garching, Germany}
\altaffiltext{2}{Kapteyn Astronomical Institute, PO BOX 800, 9700 AV Groningen, The Netherlands}
\altaffiltext{3}{Institute of Astro- and Particle Physics, Technikerstrasse 25/8, 6020 Innsbruck, Austria}
\altaffiltext{4}{MPI f\"ur extraterrestrische Physik, Postfach 1312, 85741 Garching, Germany}

\begin{abstract} 
To understand the processes that build up galaxies we investigate the stellar
structure and gas kinematics of spiral and irregular galaxies out to redshift 
1.
We target 92 galaxies in four cluster ($z=0.3$ \& 0.5) fields to study the 
environmental influence.
Their stellar masses derived from multiband VLT/FORS photometry are 
distributed around but mostly below the characteristic Schechter-fit mass.
From HST/ACS images we determine morphologies and structural parameters
like disk length, position angle and ellipticity.
Combining the spectra of three slit positions per galaxy using the MXU mode
of VLT/FORS2 we construct the two-dimensional velocity field from gas emission
lines for 16 cluster members and 33 field galaxies.
The kinematic position angle and flatness are derived by a Fourier expansion
of elliptical velocity profiles.
To trace possible interaction processes, we define three irregularity 
indicators based on an identical analysis of local galaxies from the SINGS
project.
Our distant sample displays a higher fraction of disturbed velocity fields
with varying percentages (10\%, 30\% and 70\%)
because they trace different features.
While we find far fewer candidates for major mergers than the SINS sample at
$z\sim2$, our data are sensitive enough to trace less violent processes.
Most irregular signatures are related to star formation events and less
massive disks are affected more than Milky-Way type objects.
We detect similarly high fractions of irregular objects both for the
distant field and cluster galaxies with similar distributions.
We conclude that we may witness the building-up of disk galaxies still at 
redshifts $z\sim0.5$ via minor mergers and gas accretion,
while some cluster members may additionally experience stripping, evaporation
or harassment interactions.
\end{abstract}



\section{Velocity fields of distant galaxies observed with FORS2}

Observing two-dimensional velocity fields of distant, small and faint galaxies
is still challenging and very time consuming with 3D-spectrographs even at 
large telescopes.
We took, therefore, another, more efficient approach that makes advantage of
the MXU mode of the FORS2 instrument at the VLT that allows to cut individual
slits of any orientation into a slit mask
\citep{ZKRBK09}.
For any given target, we observe three different slit positions with a central
1\arcsec-wide slit along the photometric major axis and two adjacent ones 
shifted by 1\arcsec\ along the minor axis.
Since the slit length is not limited, we cover the full extent of the galaxy 
disk along its major axis with good spatial sampling of 0.25\arcsec/pixel 
allowing to measure the rotation curve out to its flat part.
Although the three slit positions require the observation of three different 
masks, our method is still very efficient since we can target 25--30 galaxies
with the same mask.
And to achieve the necessary $S/N$ a total integration time per mask of 2.5h
is sufficient thanks to the high throughput of FORS2 and the usage of a 
VPH grism. 
The chosen grism (\textsf{600RI}) has the additional advantage that it 
provides a long wavelength range of 3000\AA.
Thus, most spectra of our targets encompass 4 (up to 7) emission lines
and many more absorption lines that can be used together to determine the
rotation of the stellar component of the galaxy.
Each emission line instead, allows to measure the kinematics of the warm gas.
A full two-dimensional velocity field (VF) for each emission line
can be constructed by a careful combination of all data points from the three 
slit positions
\citep{KZPRK08}.

\begin{figure}
\plotone{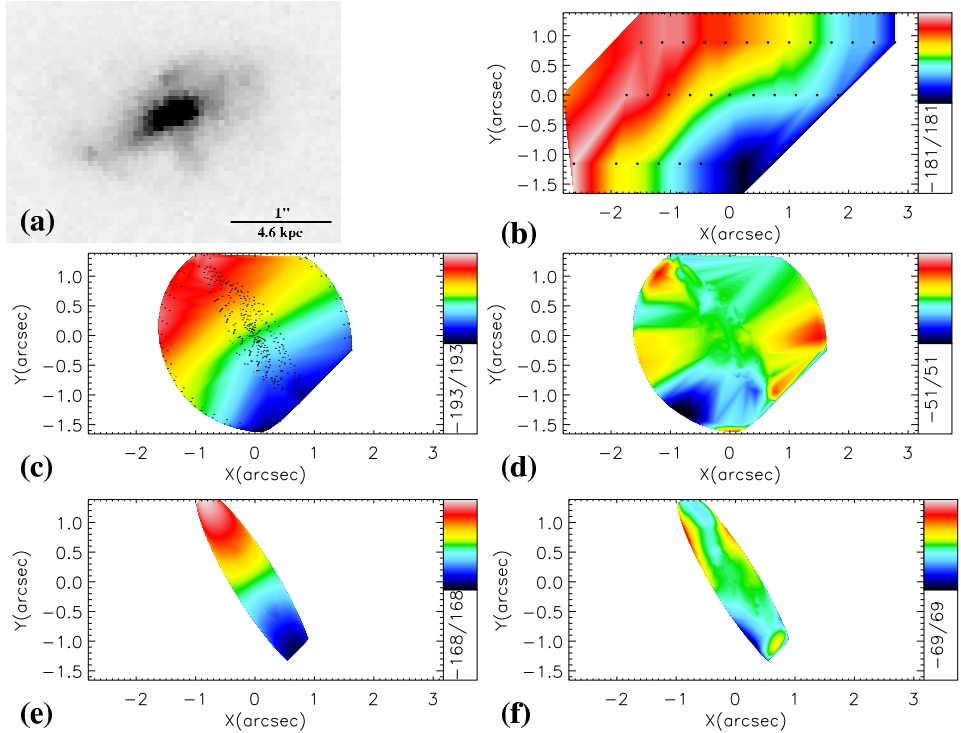}
\caption{A member of cluster MS\,2137--25 at $z=0.31$.
a) HST/ACS image in F606W filter, 
b) observed velocity field (VF) derived from the [O\,II]3727 line,
c) reconstructed VF using 6 harmonic terms, d) residual of b) and c),
e) simple rotation map, f) residual of b) and e).
See also Kutdemir et.~al (2008). }
\end{figure}

Since such a VF has an adequate spatial sampling for typical seeing and matches
the angular size of the disk, we can analyze it quantitatively with
\textsf{kinemetry}, a package originally developed for local SAURON galaxies
\citep{KCZC06}.
From fitting elliptical velocity profiles, position angles and flattenings are
found as a function of radius.
Deviations from the best fits are analyzed with a Fourier expansion, that 
traces not only the bulk motion but also indicates deviations from simple
rotation or separate kinematic components.
As an example, we show in Figure\,1  a member of a cluster at $z=0.31$
providing a thumbnail from the HST/ACS image and displaying the observed VF.
The middle panels show the reconstructed VF using a Fourier expansion to order
6 and its residual by subtracting the model from the data.
We can find the global kinematic axis by averaging the radial position angles 
and flattenings.
With these values fixed for the whole galaxy, a simple rotation map can be
created, which is shown together with its residual in the lower panels.
This galaxy has a rather smooth and regular VF, but its
(gas) kinematic axis is rotated with respect to its (stellar) photometric
axis.
Such a behavour may be caused by smooth accretion of gas infalling from 
one direction only, e.g. along a filament.

\section{Irregularities in velocity fields}

To detect irregularities quantitatively in a VF and in the same manner for all
galaxies, we define three parameters:
1.) $\sigma_{\rm PA}$: the standard deviation of the kinematic position
angles,
2.) $k_{3,5}/k_{1}$: an average value of the 3rd and 5th-order Fourier 
coefficients normalised by the rotation velocity, and
3.) $\Delta\phi$: the mean difference between photometric and kinematic
position angles across the galaxy.
While the first two trace the gas kinematics only, the last one indicates a
misalignment between the stellar and gas disks.
To find the proper value range, for which these parameters are still compatible
with undisturbed motions, we first analyzed 18 local (mostly field) galaxies 
from the SINGS sample
\citep{DCAHC06}
that have higher $S/N$ and spatial sampling.
Next, we analyze the VFs of 16 cluster
(MS1008.1-1224 $z=0.30$, 
MS2137.3-2353 $z=0.31$, 
Cl0412-65 $z=0.51$, 
MS0451.6-0305 $z=0.54$)
and 29 field ($0.1 \le z \le 1.0$) galaxies
\citep{KZPRB09}.
The fraction of irregular galaxies above the respective thresholds is lower
for the VF tracers (10\% and 30\%) than for the misalignment $\Delta\phi$
(70\%).
Their abundance and distribution is similar for both the cluster and field 
population (Fig.\,2).
While a high $\Delta\phi$ can sometimes also be caused by the presence of a
strong bar (we have 4 cases), we find correlations with recent star formation
events indicated by blue colors and clumpy substructures.
Also a trend is seen with smaller disk sizes and lower stellar masses
(which we determine from stellar population fits to our multiband photometry).

There are many more distant field galaxies with disturbed VFs than in the
local sample.
On the other hand, emission-line galaxies at $z\sim2$ from the SINS project
compatible with a major merger scenario display stronger distortions in the
Fourier analysis
\citep{SGFTB08}.
Considering all three samples, we conclude that many spiral galaxies at 
$z\sim0.5$ have not yet settled into their final configuration but are
still building-up their disks.
Since they exhibit mostly weaker irregularities, the dominant processes in
this later cosmic epoch are probably minor mergers and gas accretion.

Our cluster members also have mainly irregularities in their VFs that can be
caused by rather subtle than violent processes.
Among cluster-specific interactions, good candidates are events that strip
gas from the galaxy but leave the stellar disk with ordered motion.
Since almost all our cluster galaxies are located (in projection) within
half the cluster virial radius, those processes can be ram--pressure
stripping by the intracluster medium, thermal evaporation or harassment due to
scattering with cluster substructure and with other cluster members.
For these interactions we find similar kinematic irregularities in our 
N-body/SPH simulations
\citep{KKUSZ08,KKSZ07}.
Among those cluster members that have quite regular VFs are probably also
galaxies that are just infalling from the surrounding field, which entered
our sample due to projection.
To characterize our distant galaxies even more, we will in future also
analyze their gas metallicities, chemistry and stellar populations.

\begin{figure}
\plotone{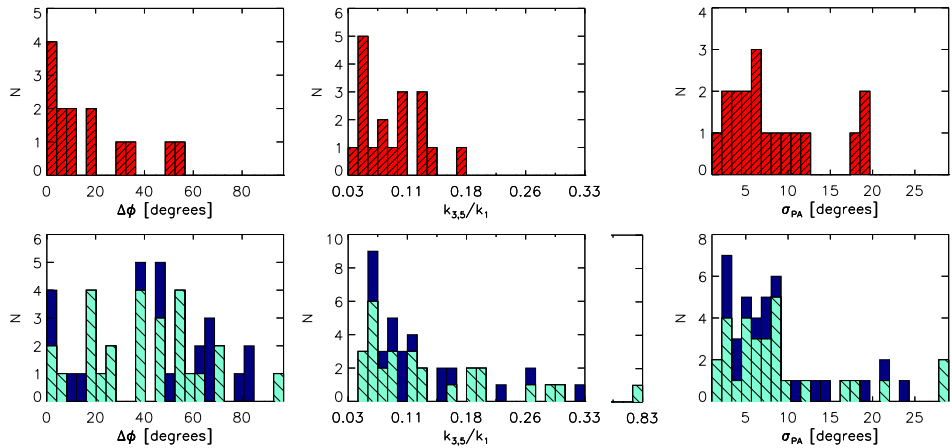}
\caption{The distribution of three irregularity parameters
for a local sample (top panels) and our distant galaxies (field: hashed,
cluster: solid histograms). 
Galaxies are classified irregular if
$\Delta\phi>25$, $k_{3,5}/ k_{1}>0.15$ or $\sigma_{PA}>20$.
See also Kutdemir et.~al (2009).}
\end{figure}

\acknowledgements 
Based on observations collected at the European Southern Observatory,
Cerro Paranal, Chile (Nos. 74.B--0592 \& 75.B--0187) 
and of the Hubble Space Telescope (No. 10635).
This work was financially supported by VolkswagenStiftung (I/76 520), 
DFG (ZI 663/6), DLR (50OR0602, 50OR0404, 50OR0301) and Kapteyn institute.


\end{document}